\documentclass[superscriptaddress]{revtex4-2}
\usepackage{graphicx}
\usepackage{dcolumn}
\usepackage{bm}
\usepackage{amsmath}
\usepackage{float}
\usepackage[colorlinks,linkcolor=blue,anchorcolor=blue,citecolor=blue,urlcolor=blue]{hyperref}

\begin{document}

\bibliographystyle{unsrt}

\title{The multi-modes Bessel-Gaussian OAM hologram encoding based on convolutional neural networks}

\author{JINJIN LI}
\affiliation{College of Physics,  Hangzhou Dianzi University, Hangzhou 310018, China}

\author{CHAOYING ZHAO}
\email{zchy49@163.com}
\affiliation{College of Physics, Hangzhou Dianzi University, Hangzhou 310018, China}
\affiliation{State Key Laboratory of Quantum Optics Technologies and Devices, Shanxi University, Taiyuan, 030006,China}
\affiliation{Zhejiang Key Laboratory of Quantum State Control and Optical Field Manipulation, Hangzhou Dianzi University, Hangzhou, 310018, China}

\begin{abstract}
\textbf{Multi-mode vortex light is a superposition of different orbital angular momentum (OAM) lights. However, as the number of OAM modes increases, the sampling constant changes. Using the traditional sparsely sampling will lead to severe loss of detail, reduced image resolution. To achieve high capacity and resolution of the OAM hologram, this paper prepares a multi-mode Bessel-Gaussian (MBG) selected hologram by stacking different mode combinations of BG phases on a MBG saved hologram in stages. Using a MBG beam with opposite combination parameters to illuminate the MBG OAM hologram, the target image can be reconstructed after the Fourier transform, and the sampling constant is flexible and controllable. The holograms encode MBG mode combination parameters. The additional degree of freedom provided by combining with MBG OAM beam offers more multiplexing channels and a higher security hologram. To further improve the quality of holograms, we first save the holograms and the corresponding MBG mode combination parameters when the quality of the obtained hologram is high based on Actor-Critic neural networks. Secondly, we gradually adjust the MBG mode combination parameters. Finally, we confirm the reasonable range of the MBG mode combination parameters.}
\end{abstract}

\maketitle

\textbf{I.Introduction}\\
The holographic display technology invented by Gabor in 1948 can reconstruct the three-dimensional object's light field wavefront through a laser beam \cite{1}. In 1966, Lohmann and Brown proposed computer-generated holography based on the sampling law and phase coding method \cite{2,3}. Optical holography includes interference recording and diffraction reconstruction and has been widely applied in fields such as three-dimensional displays \cite{4,5,6,7}, optical shaping \cite{8}, holographic multiplexing \cite{9,10}, microscopic imaging \cite{11}, optical encryption \cite{12}, and information storage \cite{13} through multiplexing in various optical dimensions such as frequency division multiplexing \cite{14}, time division multiplexing \cite{15}, polarization \cite{16}, angle \cite{17,18}, and frequency. However, the bandwidth of existing dimensions is limited.
In 1992, Allen et al. experimentally demonstrated that each photon of a Laguerre-Gaussian beam carried an orbital angular momentum (OAM) \cite{19} with a helical phase factor $e^{il\varphi}$, where $l$ is the topological charge \cite{20}. $\varphi$ is the azimuthal coordinate. Theoretically, OAM has an infinite number of orthogonal eigen-states. Therefore, utilizing OAM as a new degree of freedom for information processing can further enhance the bandwidth. With the development of nano-fabrication technologies, meta-surfaces have been considered promising devices for preparing OAM. meta-surface holography can reconstruct images with high resolution and good quality. 
 
However, due to the traditional holograms' lack of OAM selectivity, OAM has not been realized as a holographic information carrier. In July 2019, Ren et al. proposed an OAM holography scheme \cite{20}, in which OAM acts as an independent information carrier in GaN nano-pillars.

The larger the information capacity of the OAM hologram, the lower the resolution. To avoid cross-talk between the various OAM channels of the hologram, it is necessary to ensure that the OAM properties of each pixel position are not destroyed. We usually adopt the sparse sampling for the original image. However, as the number of OAM modes increases, the sampling constant increases, resulting in a severe loss of image resolution. This bottleneck problem greatly limits the improvement space for the capacity and resolution of OAM holography technology.Holograms are wavelength sensitive; the OAM beams with the same characteristics can be theoretically applied to OAM holography. The principle of the OAM holography scheme is based on the complementarity of OAM opposite angular quantum numbers and the special wavefront of OAM beams. However, since the sampling constant of the sampling matrix used in hologram production depends on the spatial frequency of the OAM beam, the number of samples of the original image decreases as the topological charge number of the OAM beam increases. When the number of samples is too small, the target image information cannot be distinguished. In order to reduce the impact of sampling constants, Ji et al. proposed the BG OAM holography scheme \cite{22} in 2024. Based on the self-healing property of the BG beam during propagation and the Fourier transform, which becomes a perfect vortex mode (different OAM modes have a fixed ring radius), the BG OAM holography scheme can effectively reduce the limitations of different OAM modes on sampling constants, improve security, and enhance anti-interference performance.

In recent years, the generation and application of multi-mode vortex light have been studied \cite{23,24}, it is a superposition state of different OAM lights. The application of multi-mode vortex light with various combined modes to holography schemes is expected to increase the number of holography multiplexing channels and improve the security of holography demodulation. The amplitude and phase of the OAM beam in the superposition state will be redistributed. The amplitude of the superposition state OAM beam shows a petal-like or multi-ring shape, while the phase is the superposition of the phases of each single-mode OAM light, which still follows the principle of complementarity of OAM opposite angular quantum numbers. Therefore, by changing the combined mode of the multi-mode vortex light, the superimposed phase can be changed to form encoded holograms with different decoding methods. So far, no scheme has been proposed to combine multi-mode vortex light with holography.

In this paper, we propose a holographic method based on multi-mode Bessel-Gauss orbital angular momentum (MBG OAM) and verify the correctness through numerical simulation. Firstly, the MBG-preserving hologram is generated by using computer-generated holography, and the phase of single-mode Bessel-Gauss beams with specific cone parameters $a$ and angular quantum numbers $l$ is superimposed to create the MBG selection hologram. When reconstructing the target image, an MBG beam with opposite selective holographic parameters ($a,l$) to the specific MBG beam is used to illuminate the MBG-OAM selection hologram, and the target image can be reconstructed after the Fourier transformation.

The designed scheme utilizes the adjustable ring radius of the BG beam to achieve OAM selective holography without destroying the OAM characteristics in higher topological cases, realizing flexible and controllable sampling constants. The encoded phase of the hologram contains multiple selective holographic parameters $(a_{n},l_{n})$ combinations, and the incident light during decoding must satisfy the superimposed state MBG beam with $(-a_{n},-l_{n})$ parameter combinations to reconstruct the image. In addition, the maximum multiplexing channel count of OAM holography is equal to the number of available OAM modes. In practice, the number of multiplexing channels is affected by the performance of the actual OAM light and spatial light modulator.\\

\textbf{II.Principles and Methods}\\

\subsection{Bessel-Gaussian beams and multi-mode vortex beams}
$\Gamma(x)$ is the gamma function, when $x$ is a positive integer
\begin{equation}
\Gamma (x)=(x-1)!
\end{equation}
$J_l(k_{r}r)$ is the first type of $l$-order Bessel functions
\begin{equation}
J_l\left(k_{r}r\right)=\sum_{m=0}^{\infty}\frac{(-1)^m}{m!\Gamma(m+l+1)}(\frac{k_{r}r}{2})^{2m+l}  
\end{equation}
In the experiment, a Bessel Gaussian beam is substituted by a Bessel beam  \cite{25}
\begin{equation}
E_{BG}\left(r,\varphi,z\right)=J_l\left(k_{r}r\right)e^{il\varphi}e^{ik_{z}z}e^{-\frac{r^2}{w_0^2}}
\end{equation}
here, $k=2\pi/\lambda$, $k_{z}=k\cos\theta$, $k_{r}=k\sin\theta$ are the wave number, the propagation direction wave number, and the radial wave number, respectively. $w_0$ is the Gaussian beam waist radius, $l$ is the topological charge of the BG beam. \\ 
Single mode vortex light has a hollow annular light field distribution.
Figure 1 shows the cross-section intensity distribution and phase distribution of different topological charges $l$.
\begin{figure}[t]
\centering
\includegraphics[width=0.6\linewidth]{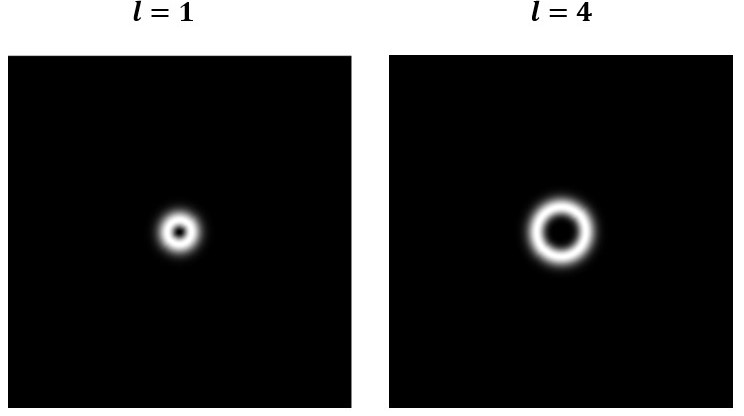}
\caption{Cross section intensity distribution of single-mode vortex beams at the waist of the beam with the different topological charges $l$.}
\end{figure}

BG beams can be generated by using axi-prisms, we have
\begin{equation}
\frac{k_r}{k} =\frac{\lambda }{a} 
\end{equation}
where $a$ is the axis prism parameters.\\

After passing through the Fourier lens, $w_a$ is the half ring width, when $r_0\gg w_a$, the BG beam becomes a perfect optical vortex (POV).
\begin{equation}
E_{POV}\left(r,\varphi\right)=i^{l-1}\frac{w_0}{w_a}e^{il\varphi}e^{-\frac{(r-r_0)^2}{w_a^2}}
\end{equation}
Actually, the ring radius of POV obtained by BG beam Fourier transform is still affected by topological charge. 
When we want to reconstruct images by using a computer-generated hologram, a single mode BG-OAM hologram without incident light undergoes a Fourier transform to obtain the target image constructed by a POV mode. 
Multi-mode vortex light contains different OAM modes, the light field distribution of the cross-section is petal shaped due to the superposition of different OAM modes, and the corresponding phase distribution is also different.
Our dielectric metasurface can possess two independent OAM modes for the PB phase that are controlled independently by their topological charges of $l_1$ and $l_2$.
\begin{equation}
E\propto e^{il_1\varphi}+e^{il_2\varphi}
\end{equation}
 The cross section component of transmitted light is expressed by Eq.(6). The corresponding light intensity distribution becomes
\begin{equation}
\left |E\right|^2 \propto \left|e^{il_1\varphi }+e^{il_2\varphi} \right |^2 =2\left \{1+\cos[(l_1-l_2)\varphi]\right\}
\end{equation} 
The amplitude of arbitrary complex electromagnetic field is only determined by $l_1$ and $l_2$. Due to the ring radii of single-mode BG beams being adjustable, we show the cross-section intensity distribution and phase distribution with the different mixed proportions of $l_1$ and $l_2$ in Fig. 2.

\begin{figure}[hp]
\centering
\includegraphics[width=0.6\linewidth]{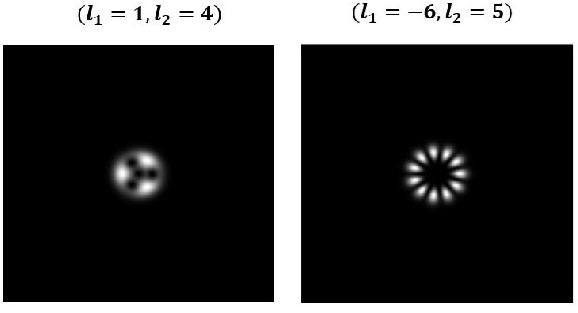}
\caption{Cross section intensity distribution of dual-mode vortex beams at the waist of the beam with the different mixed proportion of $l_1$ and $l_2$.}
\end{figure}

When the number of modes involved in stacking increases or the proportions of each mode are different, the cross-section intensity distribution and phase distribution become more and more complex.

BG beams also can be generated by using SLM loaded phase masks \cite{26}, the phase mask function is $e^{il\varphi+iar}$. The Gaussian beam is incident on the spatial light modulator(SLM) form a BG beam with radial and longitudinal components. Adjusting $a$ and $l$ can effectively control the ring radius of the generated BG vortex beam. 
The working principle of the MBG-OAM hologram is shown in Figure 3.
\begin{figure}
\centering
\includegraphics[width=0.6\linewidth]{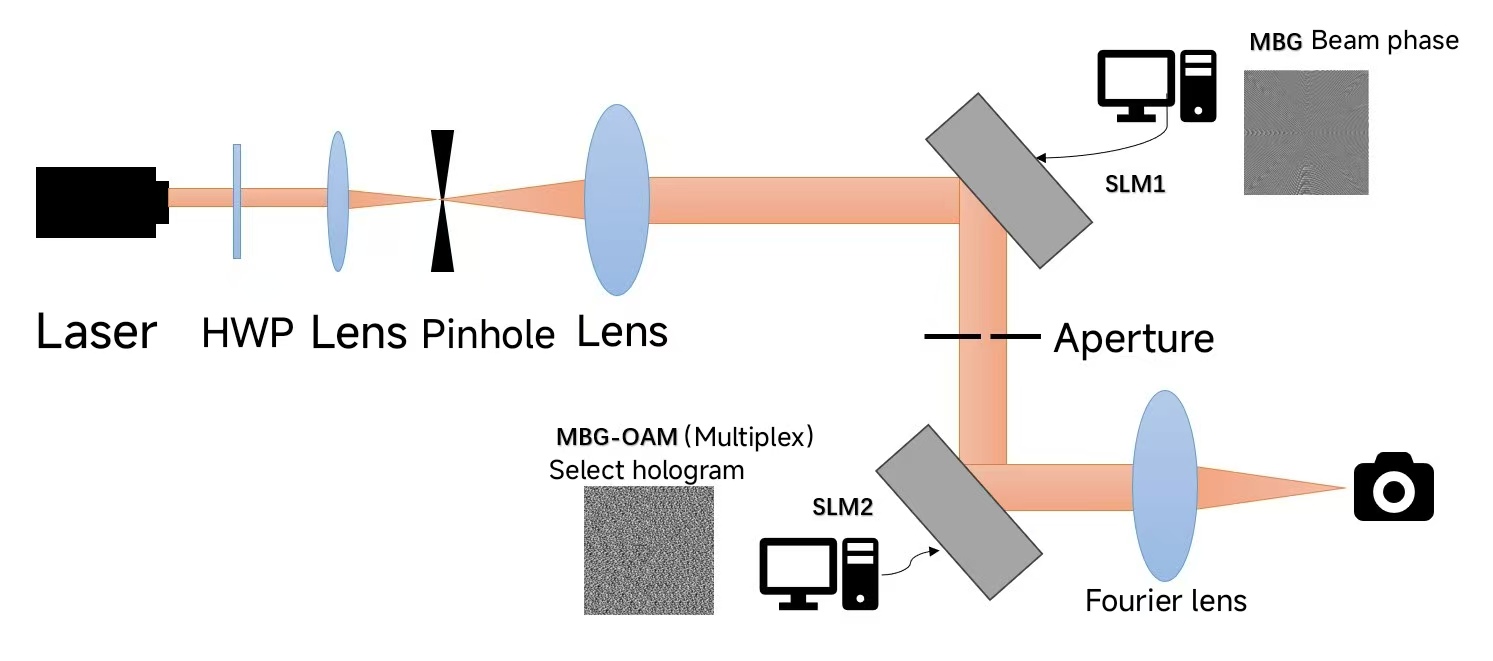}
\caption{MBG-OAM Hologram scheme.SLM: spatial light modulator. HWP:half-wave plate.}
\end{figure}

As the laser beam is modulated into polarized light parallel to the working direction of SLM1 after passing through a half wave plate, shaping and expanding through a lens group and pinholes. SLM1 is loaded with the phase of a MBG beam, so the incident polarized light can be modulated into a specific MBG beam. Only diffraction light of $l=+1$ or $l=-1$ can pass through the aperture; it is incident onto SLM2 loaded with MBG-OAM selected holograms or MBG-OAM multiplexed holograms. After passing through the Fourier lens, it is incident onto the CCD on the rear focal plane. The Gaussian point image of the target image can only be reconstructed when the phase of the MBG on SLM1 is opposite to that on SLM2.
In summary, the multi-mode vortex beams composed of BG beams can be represented as
\begin{equation}
E\propto e^{il_{1}\varphi+ia_{1}r}+e^{il_{2}\varphi+ia_{2}r}+\ldots+e^{il_{n}\varphi+ia_{n}r}
\end{equation}
Due to the ring radius of single-mode BG beams being adjustable, the radius of MBG beams is affected by the radii of each single-mode BG beam participating in the superposition.

\subsection{multi-mode Bessel-Gaussian OAM hologram}
The traditional computer-generated hologram's inability to preserve the circular features of the OAM beam is due to its continuous spatial frequency distribution. In order to achieve an OAM hologram, each sampling point needs to effectively preserve the spiral wavefront of the OAM beam. According to the opposite OAM cancellation principle, OAM-selected holograms and multiplexed holograms can be achieved.

According to the Fourier transform, the complex amplitude distribution of a beam is the spatial function and the spectral function \cite{27}
\begin{align}
\begin{split}
G(x,y)=\iint g(u,v)e^{[-i2\pi (ux+vy)]}dudv       
\end{split}
\end{align}
here, $G(x,y)$ and $g(u,v)$ represent the spatial function of the holographic plane and the spectral function of the image plane, respectively. $(x,y)$ and $(u,v)$ are the orthogonal coordinates.
In computer generated hologram, the digitized spatial function can be expressed by
\begin{align}
\begin{split}
G(x,y)=\sum_{u=1}^{M} \sum_{v=1}^{N} g(u,v)e^{[-i2\pi (ux+vy)]}      
\end{split}
\end{align}
In order to achieve OAM selected hologram, it is necessary to sample the image information with continuous spatial frequency distribution and avoid reasonably wavefront overlap, which will cause OAM beam mode characteristics damage. 

The original image sampling by a two-dimensional Dirac comb function related to the OAM spatial frequency. 
The two-dimensional Dirac comb function can be described as
\begin{align}
\begin{split}
Comb(x,y)=\sum_{m}\sum_{n}\delta [(x-md)(y-nd)]            
\end{split}
\end{align}
where $(x,y)$ is the Cartesian coordinates of the original image.  $m,n$ are the ordinal numbers of the sampling points on the $x,y$ axes, respectively.The sampling constant $d$ of the comb function is determined by the OAM spatial frequency.  We use a MBG beam; therefore, the influence of the sampling constant on the OAM hologram can be reduced by adjusting the ring radius of the BG beam. Our scheme greatly reduces the limitation of sampling constants on the hologram in high-order modes.\\
  By using the two-dimensional Dirac 
comb function $Comb(x,y)$, the sampling process of the original image $U(x,y)$ can be described as
\begin{align}
\begin{split}
D(x,y)=Comb(x,y) U(x,y)           
\end{split}
\end{align}
where the target image is $D(x,y)$,$m \times n$ is the size of the target light field.

Subsequently, a random phase $\varphi_{r}$ is superimposed on the sampled image and iterated by using the Gerchberg-Saxton(GS) algorithm \cite{28}. After reaching the preset number of iterations, the OAM saved hologram is obtained. Overlay $l$ -order BG beam phase with an axial prism parameter of $a$ on the OAM saved hologram, so that the Fourier transform of the spiral wavefront is copied to each pixel of the hologram. The  BG-OAM multiplexed hologram $D_{BG-OAM}(x,y)$ calculated according to the angular spectrum diffraction is expressed as
\begin{align}
\begin{split}
D_{BG-OAM}(x,y)=\mathcal{F}^{-1}[D(x,y)e^{2\pi i\varphi_{r}}] e^{il\varphi+iar} 
\end{split}
\end{align}
where $\mathcal{F}^{-1}$ represents the fast inverse Fourier transform. When the random phase $\varphi_r$ of the superimposed beam changes to the phase of the MBG-OAM beam, the above  equation becomes 
\begin{equation}
\begin{aligned}
&D_{MBG-OAM}(x,y) \\ &=\mathcal{F}^{-1}[D(x,y)e^{2\pi i \varphi_{r}}] (e^{il_{1}\varphi+ia_{1}r}+e^{il_{2}\varphi+ia_{2}r}+\ldots+e^{il_{k}\varphi+ia_{k}r}) \\
&=\mathcal{F}^{-1}[D(x,y)e^{2\pi i\varphi_{r})}] \sum_{k=1}^{K}e^{il_{k}\varphi+ia_{k}r} 
\end{aligned}
\end{equation}
 where $K$ is the number of the single-mode BG beams participating in superposition in the MBG beam. $l_{k}$ and $a_{k}$ are the topological charge and the axial prism parameters of the $k$ -th single-mode BG beam in the MBG beam, respectively. When the same information positions overlap, the multi-mode OAM reverts to Spin Angular Momentum(SAM) Gaussian mode superposition, which is theoretically beneficial to improve the quality of the reconstructed image. 

When we reconstruct the target image for the multiplexed hologram, based on the characteristic of restoring Gaussian mode through opposite topological charge cancellation, for example, we use $(a_1,l_1;a_2,l_2;...)$ to describe a specific MBG beam. When we reconstruct an image, theoretically, the combination parameters of the incident MBG light must satisfy $(-a_1,-l_1;-a_2,-l_2;...)$ before the image can be reconstructed. The information carried by the $k$-th hologram can be reproduced, while other MBG modes cannot recover Gaussian mode. After filtering with a comb function array, the information carried by other phase modes can be reproduced. The principle of not displaying information is described as follows
\begin{align}
\begin{split}
D_{rec_1}\left(x,y\right)=\mathcal{F}\left[D_{K-MBG-OAM}\left(x,y\right) \left(-MBG_k\right)\right] Comb\left(x,y\right)\\={S D}_{k-MBG-OAM}\left(x,y\right)
\end{split}
\end{align}
\begin{align}
\begin{split}
D_{rec_2}\left(x,y\right)=\mathcal{F}\left[D_{K-MBG-OAM}\left(x,y\right) (n_BG_k)\right] Comb\left(x,y\right)\\=n_S D_{k-MBG-OAM}\left(x,y\right)\ 
\end{split}
\end{align}
\begin{align}
\begin{split}
D_{rec_3}\left(x,y\right)=\mathcal{F}\left[D_{K-MBG-OAM}\left(x,y\right) \left(BG_k\right)\right] Comb\left(x,y\right)\\=D_{h-MBG-OAM}\left(x,y\right)
\end{split}
\end{align}
where $D_{rec}(x,y)$ represents the complex amplitude of the reconstructed image plane, $\mathcal{F}$ represents the fast Fourier transform, where the results of $D_{rec_1}\left(x,y\right)$ show the $k$-th MBG-OAM hologram.$-MBG_{k}$ represents the MBG with completely opposite combination parameters to the $k$-th MBG beam. 
The phases of the $S$ single-mode BGs carried are all restored to the point Gaussian mode. The results of $D_{rec_2}\left(x,y\right)$ show $n$ in $S$ single-mode of the $k$-th MBG-OAM hologram, OAM mode restores the Gaussian mode.$n_BG_k$ indicates that the beam is coincident with the parameter in 
$n$ single-mode BG beam collection. The results of $D_{rec_3}\left(x,y\right)$ indicate that the $h$-way MBG hologram carrying different information occurs simultaneously in image reconstruction. 
In this way, when we use MBG light with different specific combination parameters for incident multiplexed holograms, different holographic images can be reconstructed. 
In order to avoid crosstalk, we should take care to avoid information overlapping with the same encoding parameters when we use the multiplexed holograms by combining parameters of MBG light or single-mode BG light incidence.\\

\textbf{III. Simulation and Analysis}\\
Based on the computer-generated holography for target image reconstruction, the single-mode BG-OAM selected hologram without incident light converts the BG beam at each sampling point position into POV by the Fourier transform, as shown in Fig. 4.\\

\begin{figure}[t]
\centering
\includegraphics[width=0.6\linewidth]{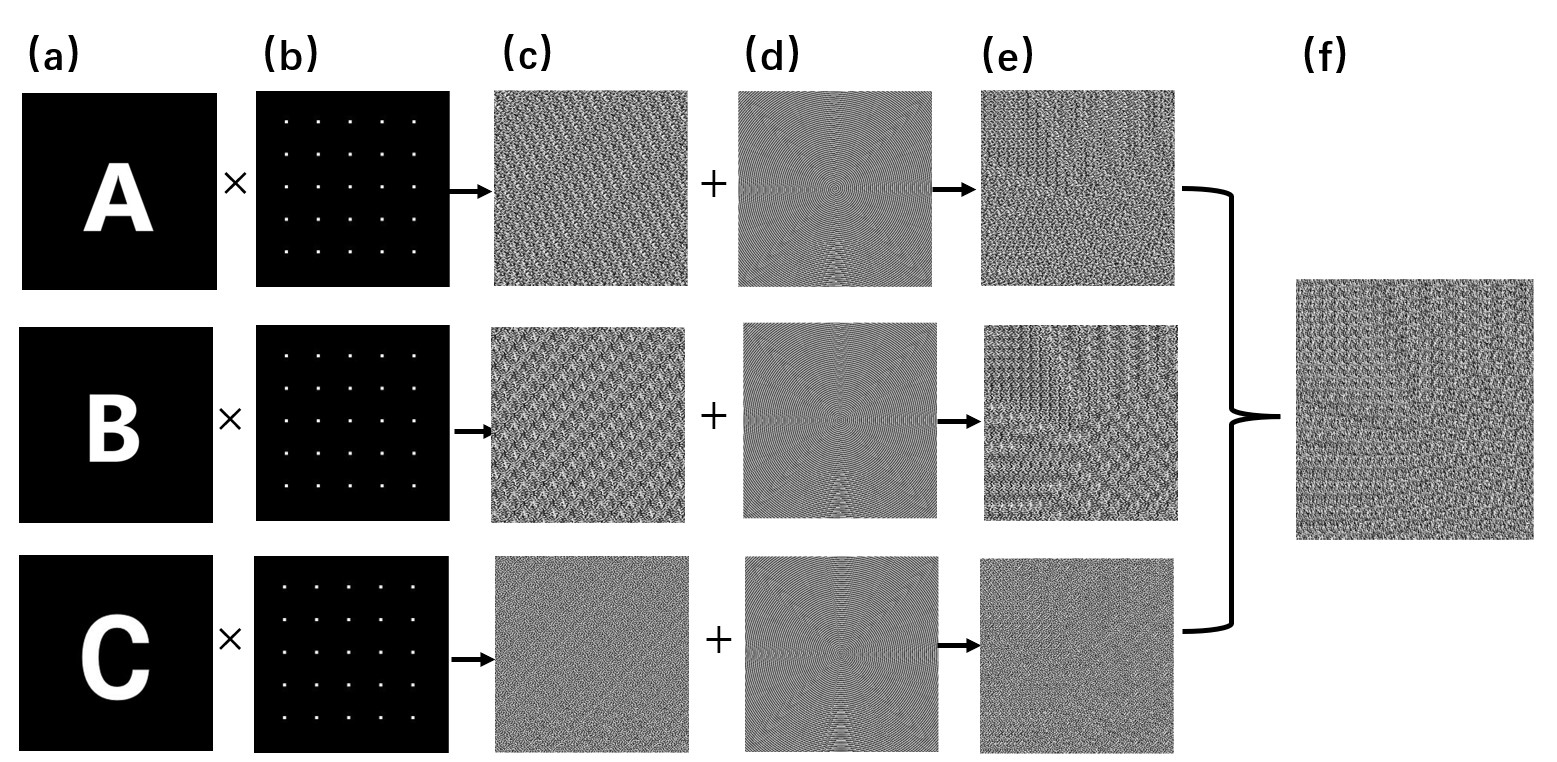}
\caption{Image reconstruction of MBG-OAM selected hologram without the incident light.}
\end{figure}

Image reconstruction of 
MBG-OAM saved holograms, MBG-OAM selected holograms, and MBG-OAM multiplexed holograms as shown in Fig. 5.\\

\begin{figure}[t]
\centering
\includegraphics[width=0.3\linewidth]{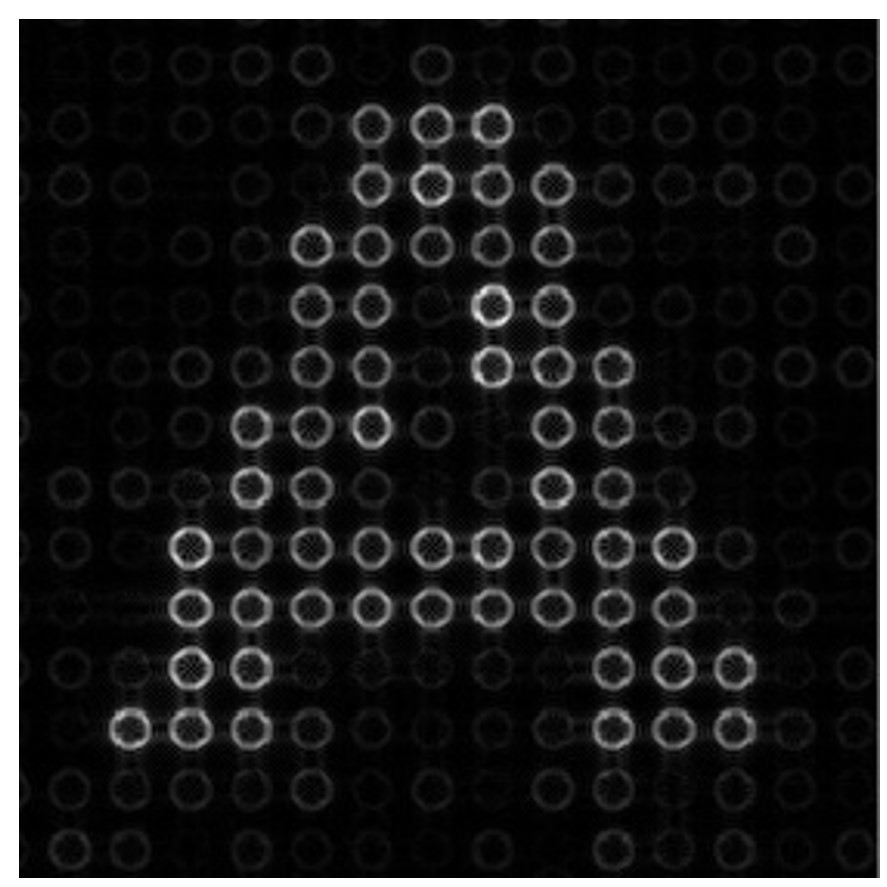}
\caption{Image reconstruction of MBG-OAM single channel multiplexing hologram:(a)original image, (b)sampling array, (c)MBG-OAM saved hologram, (d)multi-mode BG beam phase, (e)MBG-OAM selected hologram, (f)MBG-OAM multiplexed hologram.}
\end{figure}

\subsection{Multi-mode Bessel-Gaussian OAM hologram}
To verify the feasibility of the MBG-OAM holographic scheme, we conducted simulation verification based on dual-mode BG light, and the entire process was implemented by using computational holography.
The encoding parameters of the MBG loaded on SLM2 mainly include the axial prism parameters, topological charges, and the number of modes of the BG beams involved in stacking. 

Firstly, the original image "A" is sampled to obtain the corresponding MBG-OAM saved hologram, which is then overlaid with a dual-mode BG beam of $(-0.06,-3;-0.08,-2)$ to obtain the MBG-OAM selected hologram. When the incident light parameters are $(0.06,3;0.08,2)$, $(0.06,3;0.08,4)$, and $(0.06,3;0.1,2)$, respectively, the MBG-OAM selected hologram can be obtained, as shown in Fig.5.

The simulation results show that only MBG beams with $(0.06,3;0.08,2)$ can reconstruct the target image "A" when illuminating the hologram, while MBG beams with other combination parameters can choose the hologram when illuminating MBG-OAM. The filtered image appears as a circular or irregular intensity distribution, which is the superposition result of MBG beams without mutually cancelling topological charges or prism parameters. When the topological charges cancel each other, the MBG beam with the $(0.06,3;0.1,2)$ combination parameter illuminates the hologram, and after filtering, there is image information with lower brightness. This indicates that $a$ can't be used as an absolute key decoding parameter. This is because the phase of the light field is redistributed after the BG beams with different ring widths are stacked, and the intensity distribution may illuminate the original center. The quality of the reconstructed image is still affected by $a$. Therefore, it is recommended that the MBG beam used for image reconstruction be consistent with the $\left|a\right|$ of the participating BG beams.

\subsection{Multi-mode Bessel-Gaussian OAM multiplexed hologram}
When we use MBG-OAM selected holograms for multiplexed hologram, the degrees of freedom in multi-mode BG beams, such as topological charge, axial prism parameters, and the number of BG beams participating in superposition are different. The MBG-OAM selected hologram generated by different combinations carries different information, and the number of reuse channels breaks the previous limitation of only topological charge. More encoding degrees of freedom greatly improve the number of multiplexing channels and decoding security.
In order to verify the feasibility of the MBG-OAM multiplexed hologram, we conduct simulation verification based on dual channel MBG-OAM hologram. Firstly, the original image "A" is sampled to obtain the corresponding MBG-OAM saved hologram, which is overlaid with a dual-mode BG beam with a combination parameter of $(-0.06,-3;-0.06,-2)$ to obtain the MBG-OAM selected hologram. (1) Sample the original image "B" to obtain the corresponding MBG-OAM saved hologram, and overlay it with a dual-mode BG beam with a combination parameter of $(-0.1,-8;-0.1,-10)$ to obtain the MBG-OAM selected hologram. (2) Overlay MBG-OAM selected hologram 1 and MBG-OAM selected hologram 2 to create an MBG-OAM multiplexed hologram, with overlapping positions of the two target images in the hologram. According to the variable method, MBG optical incident multiplexed holograms were tested and simulated by changing the topological charge, the topological charge, and the parameters of the prism, and the parameters of the prism. As follows, when the combination parameters of the incident MBG light are $(0.06,3;0.06,2)$, $(0.06,3;0.06,2)$, and $(0.1,5; 0.02,20)$, the MBG-OAM multiplexed hologram is irradiated to obtain in Fig.5.
\begin{figure}[t]
\centering
\includegraphics[width=0.6\linewidth]{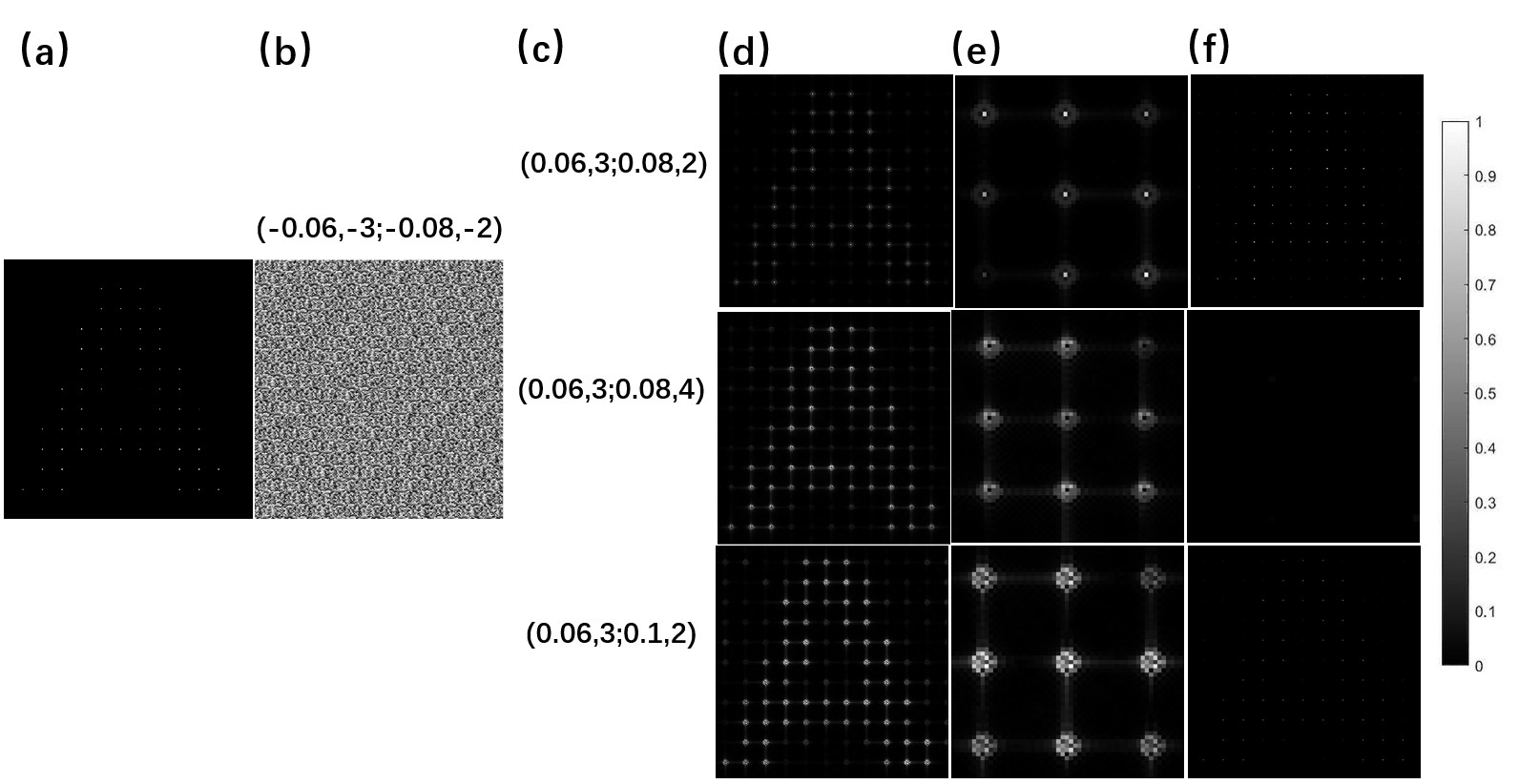}
\caption{Image reconstruction of MBG-OAM dual channel multiplexed hologram: (a)target image, (b)MBG-OAM multiplexed hologram and parameter combination, (c)combination parameters of incident MBG beam, (d)unfiltered reconstructed image, (e)details unfiltered reconstructed image, (f)filtered reconstructed image.
}
\end{figure}

Figure 6 shows that the corresponding target image "A" and target image "B" can be reconstructed when the combination parameters of the incident MBG beam are $(0.06,3;0.08,2)$ and $(0.1,5;0.12,1)$, respectively. Otherwise, due to the decoding parameter $(0.06,3;0.12,1)$ and the combination parameters $(-0.1,-5;-0.12,-1)$ or $(-0.06,-3;-0.08,-2)$ that
cannot cancel each other, the reconstruction fails. The results indicate that the MBG-OAM multiplexed hologram is feasible.
\begin{figure}[t]
\centering
\includegraphics[width=0.6\linewidth]{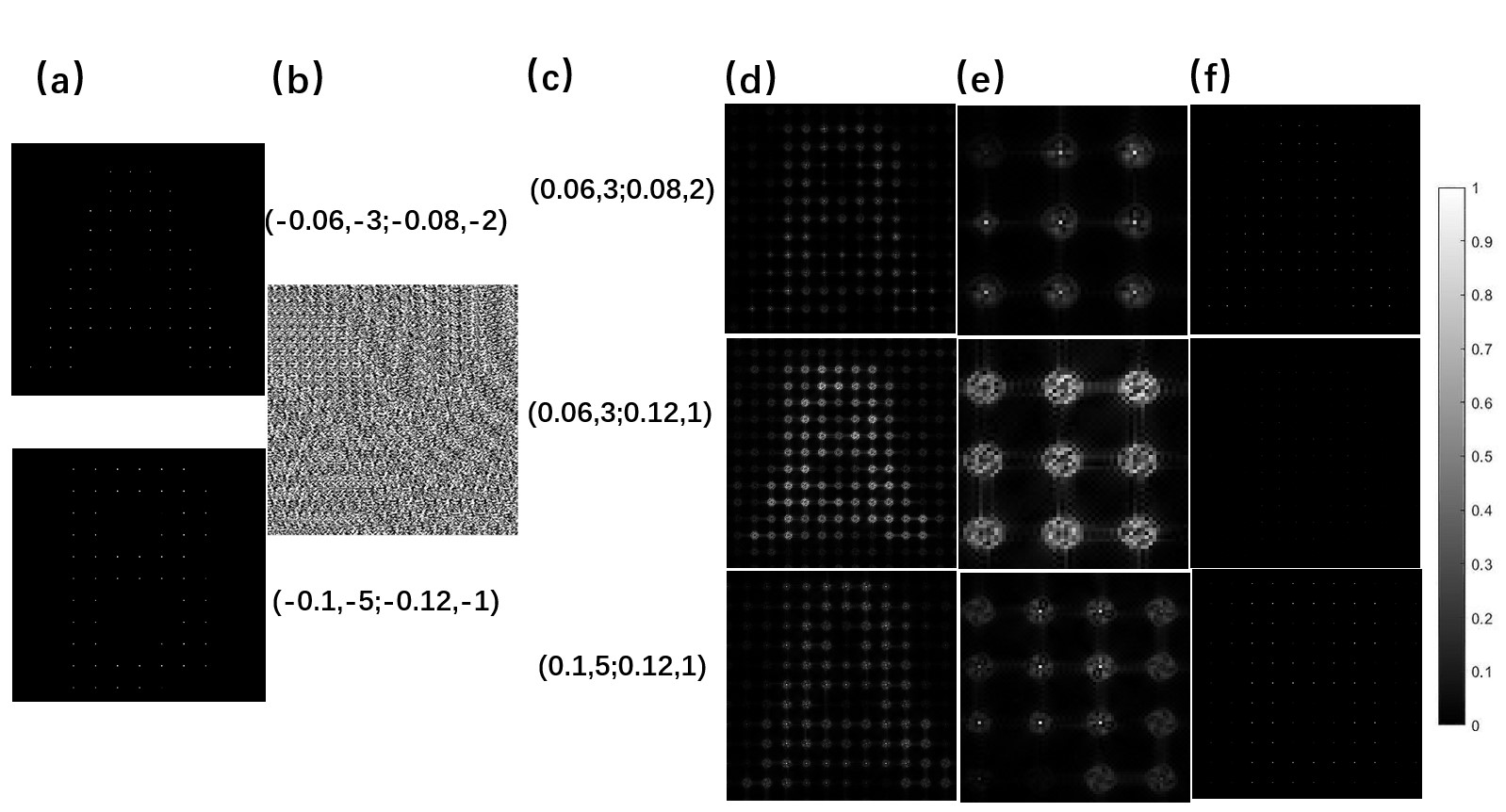}
\caption{Image reconstruction of MBG-OAM dual-channel multiplexed holograms:(a)pre-sampling image information,(b)MBG-OAM multiplexed holograms and parameter combinations,(c)Combination parameters of the incident MBG beam,(d)unfiltered reconstructed image,(e)filtered reconstructed image.}
\end{figure}
It shows that when the incident MBG beam and BG beam combination parameters are taken as $(0.02,18)$, $(0.02,20)$, $(0.06,3; 0.06,2)$,
$(0.1,10)$, C, CD, A, B information is reconstructed, and the above simulation results show that we propose an MBG-OAM multiplexed hologram. The different information hologram can encode the same phase, and selective reconstruction of a single message or multiple messages can be implemented. Figure 8(B),(E) is a detail and Figure 6(b),(e) is a partial detail, respectively. In Figure 7(a), the positions of the four non-overlapping messages are labeled as 1234, as shown in Figure 8(A), and in Figure 7(e1), the information "C" at position 3.
The filtered reconstruction is shown in Figure 8 (E1), while the "A", "B", "D" at the position 124 cannot be reconstructed after filtering. Figure 8(E2),(E3) shows the figure detail of position 34 corresponds to  Figure 7(e2), detail of position 1 is shown in Figure 8(E4) corresponds to Figure 7(e3), and location 5 of Figure 7(e4) corresponds to Figure 8(E5). Except for the above positions, there is no information reconstruction after filtering the rest of the positions.
\begin{figure}[t]
\centering
\includegraphics[width=0.6\linewidth]{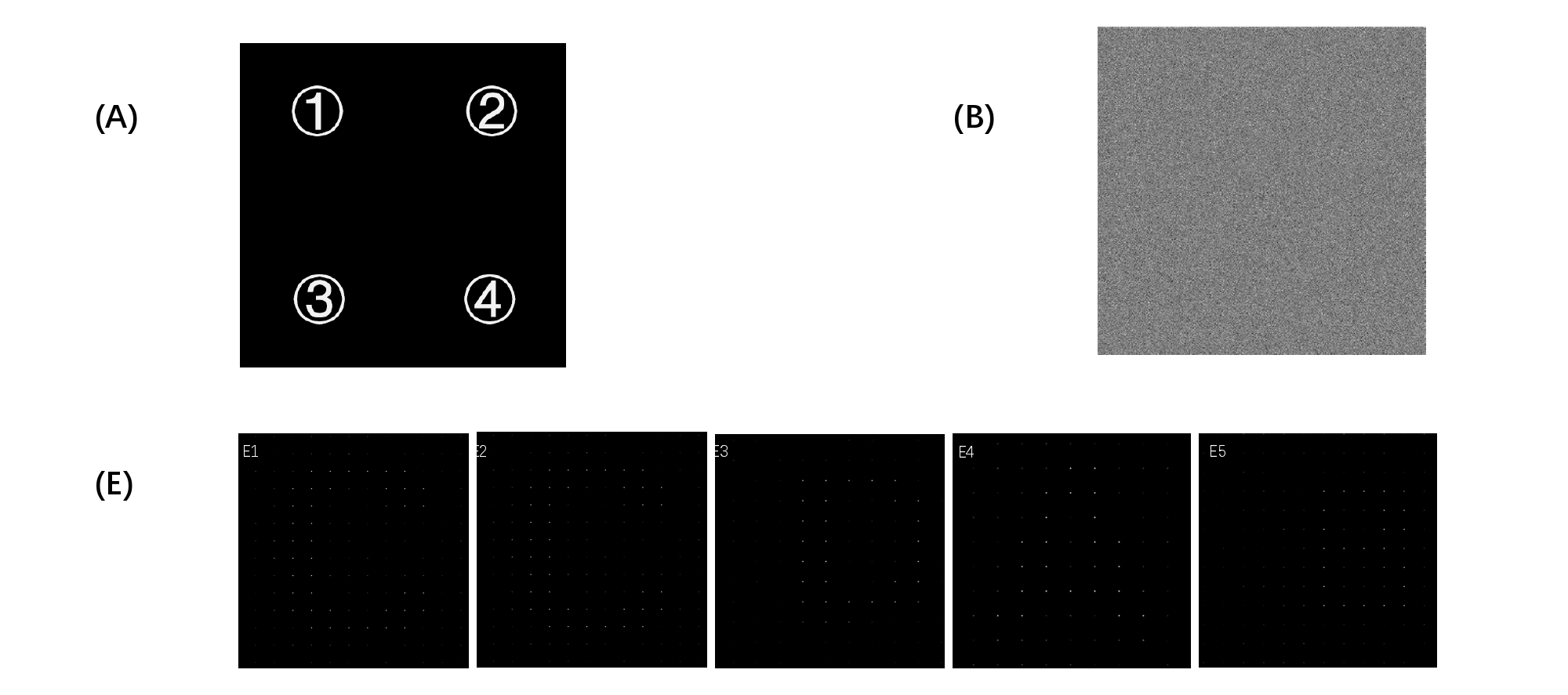}
\caption{Image reconstruction of MBG-OAM dual channel multi-
plexed hologram:(A) Image information location marker map, (B) MBG-OAM multiplexed hologram, (E) filtered information location detail map.}
\end{figure}
Because the single-channel simulation target image is a position overlap multiplexing situation, there is a serious crosstalk problem, so it is possible to reduce crosstalk by multiplexing different positions or the frequency shift method \cite{13}. The frequency shift method can break through the number of SLM pixels and achieve large-scale image solutions Encode. Crosstalk can also be effectively reduced by increasing the difference in the multiplexed topological load. Improvements to the topological load between multiple modes in a single image difference can also reduce crosstalk. Since the Fourier transform is required to reconstruct the image, it is easy to lead to small topological charges in this process.
The range cannot be distinguished, resulting in the situation that the Gaussian mode of the opposite parameter is not recovered, so the value of the parameter needs to be reconsidered. In addition, 
the gray layer and the imperfect calibration of the SLM can lead to noise and loss.  In order to assert the result, we use Mean Square Error (MSE), Peak Signal-to-Noise Ratio (PSNR), and Structural Similarity Index (SSIM)to measure the differences, as shown in Table 1; we give out a measure of the target image and the reconstructed image to calculate the results.\\

\begin{table}
\caption{\label{tab:table1}Using three indices to evaluate the calculation results of a target image and a reconstructed image.}
\begin{tabular}{|c|c|c|c|}
\hline
& $A(0.06,3;0.08,2)$ & $A(0.06,3)$ & $A(0.08,2)$ \\\hline
MSE & $2.7031^{-5}$ & $3.5985^{-5}$ & $3.0477^{-5}$\\\hline
PSNR &93.8121 &92.5695 & 93.2911 \\\hline 
SSIN &0.9867 &0.9905 & 0.9938 \\\hline 
\end{tabular}
\begin{tabular}{|c|c|c|c|}
\hline
&$C(0.02,18)$ &$A(0.06,3;0.06,2)$ &$B(0.1,10)$ \\\hline
MSE &$9.1443^{-5}$ &$5.7059^{-6}$ &$5.6846^{-6}$\\\hline
PSNR &98.5193 &100.5676 &100.5838\\\hline 
SSIN &0.9937 &0.9963 &0.9852 \\\hline 
\end{tabular}
\begin{tabular}{|c|c|c|c|}
\hline
&$C(0.02,15;0.02,18;0.02,20)$ &$B(0.1,8;0.1,10)$ \\\hline
MSE &$2.3413^{-5}$ &$5.3375^{-6}$ \\\hline
PSNR &98.4928 &100.8574 \\\hline 
SSIN &0.9938 &0.9963 \\\hline 
\end{tabular}
\end{table}

\textbf{III. Convolutional neural networks}\\
In order to reduce the computational complexity, after every two convolution layers, we add a max pooling layer to reduce the amount of data in the characteristic graph. Feature extraction from images through four Convolutional layers, the feature quantity of the final fully connected layer reaches to $262144$. The dropout layer can randomly delete some cells to reduce the complex co-adaptation relationship between neurons. Therefore, a hidden layer neuron cannot rely on other specific neurons to correct its errors, and forces the network to learn more robust features. After two fully connected layers, an activation function, and a dropout layer, four classifications can be obtained.

This study proposes an automated optimization method for hologram generation parameters based on a deep reinforcement learning framework. By designing the Actor-Critic dual network architecture, the DDPG intelligences take the hologram parameter matrix as the input state, and dynamically generate four key parameters, namely, topological charge $l$ , axial prism parameter $a$ , sampling parameter $T$ , and iteration number $n$, where $a$ determines the beam radius of the OAM beam, and $T$ determines the distribution density of the sampling points. The Actor network outputs a continuous four-dimensional action vector after extracting the features, which is mapped into integer values or decimal values that meet the physical constraints to ensure the feasibility of the parameters. After processing and mapping, it will be turned into integer or decimal values that meet the physical constraints to ensure the feasibility of the parameters; Critic network can evaluate the value of the action, calculate the quality of the hologram image, and guide the optimization direction of the strategy according to the quality score obtained. The reward function integrally calculates the Structural Similarity (SSIM) and Peak Signal-to-Noise Ratio (PSNR), assigns 60\% and 40\% weights, respectively, and sets the corresponding reference values to make the data more reliable. And the exploration noise mechanism is introduced in the training, which encourages extensive exploration through high noise in the initial stage and gradually attenuates the noise in the later stage to stabilize the strategy, and combines the experience playback and soft update technique of the target network to improve the training stability and convergence efficiency effectively, and finally outputs high-quality holograms and the corresponding optimal parameter combinations.

In the network architecture of this study, the generation parameters of holograms are outputted through the Actor network, and the quality of holograms generated by these parameters is evaluated through the Critic network, and then the parameters of the Actor network are updated according to the policy gradient.

Its input layer directly receives image data as a 612×612×3 random noise image, which is used to simulate the state space. Next, feature extraction is performed in the convolutional layers, the first layer uses 8×8 convolutional kernels, 32 filters, and a step size of 4. The second layer uses 4×4 convolutional kernels, 64 filters, and a step size of 2. Each convolutional layer is followed by an activation layer, which increases the nonlinear expressive capability. After passing through the fully connected layers, a four-dimensional action vector is generated, which corresponds to the normalized values of the four parameters $l$,$a$,$T$,$n$.
	
Its input layer is divided into state input and action input, the state input is the same as the Actor network and receives image data as a 612×612×3 random noise image; the action input is to receive four-dimensional action vectors, which are input separately from the state. State feature extraction is performed, and image features are extracted by the same independent convolutional layer as the Actor network, and features are compressed using a fully connected layer. The merging layer will merge the compressed state features with the four-dimensional action vectors, and the merged features are processed through the fully connected layer, and after connecting the activation layer, a scalar Q-value is output, which represents the value of the state-action pair.
	
The first is action generation, which adds exploratory noise that can dynamically adjust the noise intensity to the actions output from the Actor network, and the noise intensity decays by an exponential law.

\begin{equation}
\begin{split}
\sigma_t=0.6\cdot e^{-episode/(maxEpisodes/3)}+0.05
\end{split}
\end{equation}
where $maxEpisodes$ is the maximum number of training rounds (set to 60). At the beginning of training, the algorithm enhances exploration by introducing exploration noise. For example, when generating the topological charge $l$, the initial noise amplitude is set to 0.65, which makes the parameter may break through the preset range [-5,5] and try unconventional values such as -6 or 7. As the number of training rounds advances, the noise intensity decays exponentially, and this decay mechanism makes the first 20 rounds focus on global search, and then the last 40 rounds progressively lock the high-scoring region, and the noise exponent decays to 0.05, which is gradually biased towards the Utilization.

Next is the parameter mapping, from the normalized values of the four parameters generated by the Actor network, the final output action needs to be mapped to a specific parameter range, where the topological charge $l$ ranges from [-5,5], which can be set to one decimal due to its physical continuity; the axial prism parameter $a$ ranges from [0.05,0.07], three decimal; the sampling constant $T$ is an integer with an interval of  [10,30]; the number of iterations $n$ is an integer with an interval of [1,20].

The original pattern can be computed to get the OAM preservation hologram after sampling array, and then superimpose the spiral phase of a certain order on the preservation hologram to generate the OAM selection hologram. After that, it is necessary to score the quality of the hologram and make the strategy of the next action in combination with the reward function, so the design of the reward function directly affects the convergence direction of the algorithm. In this study, the combined score of Structural Similarity (SSIM) and Peak Signal-to-Noise Ratio (PSNR) is used as the reward signal, and when there is an improvement in the score, a positive reward is obtained, and the rest is a negative reward. The scoring function is realized by calculating the SSIM and PSNR of the reconstructed and sampled images respectively, and then fusing the two 6:4 weighted for the combined score, which can be expressed by the formula.

\begin{equation}
\begin{split}
score=0.6\cdot SSIM+0.4\cdot(PSNR/60)
\end{split}
\end{equation}

In the SSIM formulation, we use the subscript $x$ to denote the reference image, i.e., the value corresponding to the sampled image, which serves as the benchmark for quality assessment, and the subscript $y$ to denote the image to be assessed, i.e., the value corresponding to the reconstructed image, which serves as the target for similarity comparison.

SSIM is an image quality assessment method based on the characteristics of the human visual system, which can be used to measure the realism of model-generated images by quantifying the perceived similarity of two images in multiple dimensions. By evaluating the similarity of brightness, contrast and structural information between the reconstructed image and the sampled image, which is more in line with human subjective perception, 
\begin{equation}
\begin{split}
SSIM=\dfrac{(2\mu_x\mu_y+C_1)(2\sigma_{xy}+C_2)}{(\mu_x^2+\mu_y^2+C_1)(\sigma_x^2+\sigma_y^2+C_2)}
\end{split}
\end{equation}

Where $\mu$ is the mean value, which is used to measure luminance; $\sigma$ is the standard deviation, which is used to measure contrast; $\sigma_{xy}$ is the covariance of the two images, which is used to measure the structure; and $C_1, C_2$ is a constant, which is used to prevent computational errors when the denominator is zero. In the evaluation, the closer the value of SSIM is to 1, that is, the higher the structural similarity.

PSNR is an image quality evaluation metric based on pixel-level numerical differences, which is mainly used to quantify the degree of distortion between reconstructed and sampled images. Its calculation is based on the mean square error (MSE), which is calculated by comparing the difference in the luminance value of each pixel point of two images with the formula:

\begin{equation}
\begin{split}
\begin{aligned}
MSE = \frac{1}{M \cdot N} \sum_{i=1}^{M} \sum_{j=1}^{N} \left[ I(i,j) - K(i,j) \right]^2\\
PSNR=10\cdot\log_{10}(\frac{MAX_I^2}{MSE})
\end{aligned}
\end{split}
\end{equation}

In mean square error $MSE$, $I(i,j)$ is the gray value of the sampled image at pixel position $(i,j)$; $K(i,j)$ is the gray value of the reconstructed image at pixel position $(i,j)$; and $M\cdot N$ is the total number of pixels of the image with dimensions $M$ in width and $N$ in height. In PSNR calculation, $MAX_I$ is the maximum possible value of a pixel, which depends on the bit depth of the image, and in 8-bit image, $MAX_I=255$.The higher value of PSNR indicates the better quality of reconstruction. In the simulation calculation, the reference value of 60 dB is set because it is observed that the PSNR obtained from the simulation is between 50 dB and 55 dB, and the problem of the magnitude difference is solved by normalizing the parameters to make the calculation of the scoring function more convenient.

The training loop of the simulation is constructed based on the Actor-Critic algorithm with DDPG intelligences, generating the hologram parameters $l,a,T,n$ through the Actor network, combining with exploratory noise, the dynamic boundary adjustment module to screen the historical high scoring samples every 5 rounds, and updating the parameter ranges after expanding the buffer interval by $10\%$ in order to focus on the high-quality regions; and screen out the samples with quality ratings in the top $20\%$ to ensure that the parameter range adjustment is based on high-quality parameter combinations only. the Critic network evaluates the quality of the parameters. The Critic network evaluates parameter quality and calculates a comprehensive score based on $60\%$ of SSIM and $40\%$ of PSNR to drive strategy optimization. The experience playback pool stores state-action-reward tuples, soft updates the target network with a smoothing factor coefficient $\tau=0.001$, performs multiple rounds of stable training, and ultimately terminates when the target score is reached or the maximum number of rounds is 60, saves the best-quality holograms and corresponding parameters, and comparisons of reconstructed and sampled images to a .mat file, which can be read at any time by the inference script for the scored highest result.

During the training loop, the dynamic parameter range adjustment mechanism and the comprehensive quality scoring mechanism of the hologram show significant synergistic evolutionary utility. In this experiment, two schemes are mainly designed for the superposition mechanism of spiral phase for comparison: one is to superimpose the spiral phase field of a single topological charge $l$ to realize the hologram generation; the other adopts the method of superposing the double spiral phase of $l_1$-order and $l_2$-order to generate holograms, and constructs the multi-dimensional light field regulation system to make the experimental data more diversified. The original image used in this simulation is the capital letter A.

\subsubsection{Single spiral phase scheme}
As shown in Table 2, by parsing the optimal parameter combinations, holograms, and reconstructed images compared with sampled images for each time, we can obtain Fig. In the first training loop, the optimal parameter combination is: $l=-2.0$, $a=0.054$, $T=28$, $n=1$, corresponding to a PSNR of $51.99dB$, a SSIM of $0.9952$, and a composite image quality score of $0.9438$. In the second training loop, the optimal parameter combination is: $l=2.7$, $a=0.050$, $T=29$, and $n=1$. $T=29$, $n=1$, corresponding to a PSNR of $52.60dB$, an SSIM of $0.9955$, and a composite image quality score of $0.9479$.

\begin{table}[!ht] 
	\centering 
	\caption{Single Spiral Phase Training Loop} 
	\label{tab2} 
	\scalebox{1.3}{
    \begin{tabular}{c|cccc|c} 
		\hline 
		R &$l$ &$a$ &$T$ &$n$ &Scores \\  
		\hline
		1 & [-5.0,5.0] &[0.050,0.070] & [10,30] &[1,20] & 0.9392\\
		5 & [-1.7,3.3] &[0.050,0.070] & [23,30] &[1,2] & 0.9434\\
		10 & [1.9,3.3] &[0.050,0.055] & [27,30] &[1,2] & 0.9434\\
		60 & [2.0,3.3] &[0.050,0.055] & [27,30] &[1,2] & 0.9479\\
		\hline
	\end{tabular}
    }
\end{table}

The experimental results show that under the single-spiral phase scheme, the parameter ranges quickly converge to the local optimal region through the dynamic boundary adjustment mechanism, and the two independent training loops achieve the highest composite scores of 0.9438 and 0.9479, respectively, in which the positive and negative preference differences of the topological constants $l$ reveal the potential influence of the directions of the positive and the negative phases in the phase modulation on the image quality.

The double spiral phase scheme by introducing the superposition mechanism of the composite phase, verifying the gain effect of multidimensional optical field modulation on holographic imaging. The strategy of parameter dynamic tuning under this scheme also shows remarkable sensitivity, with the high convergence of the topological constant $l_2$ and the sustained high locking of the sampling parameter $T$, which reflect the key roles of the spiral phase dominance and the high-frequency information retention, respectively. 

Two independent training loops show more complex light field optimization properties through multidimensional parameter co-modulation as shown in Table 3. By parsing the optimal parameter combinations, holograms, and reconstructed images compared with the sampled images each time, we can obtain hologram Figure. In the first training loop, the optimal parameters $l_1=0.7$, $l_2=4.8$, $a=0.056$, $T=29$, $n=1$ corresponds to a PSNR of $52.43dB$, an SSIM of $0.9962$, and an overall image quality score of $0.9472$. In the second training loop, the optimal parameters $l_1=2.0$, $l_2=4.8$, $a=0.063$, $T=29$, $n=1$ corresponds to a PSNR of $52.61dB$, a SSIM of $0.9957$, and a composite image quality score of $0.9482$.

\begin{table}[t] 
\centering 
\caption{Double Spiral Phase Training Loop} 
\label{tab4} 
\scalebox{1.3}{
\begin{tabular}{c|ccccc|c} 
\hline 
R &$l_1$ &$l_2$ &$a$ &$T$ &$n$ & Scores \\  
\hline
1&[-5.0,5.0] &[-5.0,5.0] &[0.050,0.070] &[10,30] &[1,20] &0.8975\\
5&[-3.1,3.0] &[1.7,5.0] &[0.063,0.070] &[21,30] &[1,3] &0.9377\\
10&[-3.1,3.0] &[3.2,5.0] &[0.063,0.070] &[23,30] &[1,2] &0.9386\\
60&[0.5,3.0] &[3.7,5.0] &[0.063,0.070] &[25,30] &[1,2] &0.9482\\
\hline
\end{tabular}
}
\end{table}
The effectiveness of the dynamic optimization mechanism of hologram parameters based on the Actor-Critic deep reinforcement learning framework and the reasonableness of the image quality scoring mechanism are verified through the comparative simulation experiments of the single-spiral phase and double-spiral phase schemes.
 Analysis the training results under the triple-spiral phase scheme shows that we can achieve a significant improvement over the double-spiral phase scheme. The multiple-spiral phase superposition scenario does improve the holograms quality to some extent as shown in Table 4. The parameters $l_1=-3.7$, $l_2=1.6$, $l_3=3.0$, $a=0.057$, $T=29$, $n=1$ corresponds to a PSNR of $52.74dB$, an SSIM of $0.9960$ and a composite image quality score of $0.9492$.

\begin{table}[!ht] 
\centering 
\caption{Triple Spiral Phase Training Loop} 
\label{tab5} 
\scalebox{1.3}{
\begin{tabular}{c|cccccc|c} 
\hline 
R&$l_1$ &$l_2$ &$l_3$ &$a$ &$T$ &$n$ & Scores \\  
\hline
1&[-5.0,5.0] &[-5.0,5.0] &[-5.0,5.0] &[0.050,0.070] &[10,30] &[1,20] &0.9277\\
5&[-5.0,5.0] &[1.4,1.8] &[2.6,3.2] &[0.053,0.065] &[27,30] &[1,2] &0.9416\\
10&[-3.8,-2.1] &[1.4,1.8] &[2.6,3.2] &[0.053,0.065] &[27,30] &[1,2] &0.9488\\
60&[-3.8,-3.4] &[1.4,1.8] &[2.6,3.2] &[0.053,0.065] &[27,30] &[1,2] &0.9492\\
\hline
\end{tabular}
}
\end{table}

The dynamic parameter range adjustment mechanism and the change in quality scores showed significant differences. Overall, the parameter dynamic tuning strategy achieves progressive optimization of the scoring system through stage contraction with directional offset, but the sensitivity difference of different parameter dimensions significantly affects the final convergence characteristics.

Overall, the sampling constant $T$ in both schemes are selected higher parameter range, the surface of the high sampling constant corresponds to the hologram generation process can be more retained information to enhance the quality of holograms; the number of iterations $n$ in the two schemes are fast convergence to the minimum value of the interval, it can be analyzed that its gain effect on the quality of the image is very limited.

In addition, the experiments also designed and verified the synergy between exploring the noise attenuation mechanism and the design of the empirical playback pool, and achieved a stable improvement of the comprehensive score from the initial 0.8975 to 0.9482 within 60 rounds by balancing the global search and local optimization. All these results and analyses provide simulation support for the intelligent optimization of high-dimensional light field parameters in complex scenes, and also point out the direction for the improvement of the subsequent algorithms and adaptive boundary adjustment mechanisms.

\textbf{IV. Conclusion}

In order to improve the limitation of the number of channels for single-mode multiplexed hologram, based on the principle of multi-mode vortex beam superposition state and opposite OAM cancellation to recover Gaussian mode, combined with OAM hologram, we propose a new scheme to produce multi-mode Bessel Gaussian selected holograms and multiplexed holograms by sequentially stacking different parameter combinations of Bessel Gaussian phase on multi-mode Bessel Gaussian saved holograms. Using MBG beams with completely opposite parameter combinations to illuminate MBG-OAM holograms, the target image can be reconstructed through the Fourier transform. This scheme retains the advantages of BG-OAM hologram, with flexible and controllable sampling constants, and increases the encoding freedom of the hologram. When decoding the incident light, the corresponding mode combination parameters must be met simultaneously to reconstruct the image, effectively improving the security of OAM holography and the number of multiplexing channels.

\textbf{Acknowledgements}
We would like to express my sincere gratitude to Xufeng Yuan, a graduate student and Zhang Wuwei, an undergraduate student, who have already graduated, for assisting me to process some figures before they graduated. 

\textbf{Funding.} 
Financial support from the project funded by the State Key Laboratory of Quantum Optics Technologies and Devices, Shanxi University, Taiyuan, China(Grants No.KF202503).\\

\textbf{Disclosures.} The authors declare no conflicts of interest.\\

\textbf{Data availability statement.} Data underlying the results presented in this paper are not publicly available at this time but may be obtained from the authors upon reasonable request.\\

\textbf{Declaration of Competing Interest}\\
The authors declare that they have no known competing financial interests or personal relationships that could \\ have appeared to influence the work reported in this paper.

\bibliography{main}

\end{document}